# Orbital-resolved single atom magnetism measured with X-ray absorption spectroscopy

Aparajita Singha[1,2,†], Daria Sostina[1,2,†], Christoph Wolf[1,2,†], Safa L. Ahmed[1,3], Denis Krylov[1,2], Luciano Colazzo[1,2], Pierluigi Gargiani[4], Stefano Agrestini[4], Woo-Suk Noh[5], Jae-Hoon Park[5,6], Marina Pivetta[7], Stefano Rusponi[7], Harald Brune[7], Andreas J. Heinrich[1,3], Alessandro Barla[8], and Fabio Donati[1,3,*]

[1]*Center for Quantum Nanoscience, Institute for Basic Science (IBS), Seoul 03760, Republic of Korea*
[2]*Ewha Womans University, Seoul 03760, Republic of Korea*
[3]*Department of Physics, Ewha Womans University, Seoul 03760, Republic of Korea*
[4]*ALBA Synchrotron Light Source, Cerdanyola del Vallès, 08290, Catalonia, Spain*
[5]*MPPC-CPM, Max Planck POSTECH/Korea Research Initiative, Pohang 37673, Korea*
[6]*Department of Physics, Pohang University of Science and Technology (POSTECH), Pohang 37673, Korea*
[7]*Institute of Physics, École Polytechnique Fédérale de Lausanne, Station 3, CH-1015 Lausanne, Switzerland*
[8]*Istituto di Struttura della Materia (ISM), Consiglio Nazionale delle Ricerche (CNR), I-34149 Trieste, Italy*

[†] *Equally contributing authors*
[*]*E-mail: donati.fabio@qns.science*

## Abstract

Lanthanide atoms and molecules are promising candidates for atomic data storage and quantum logic due to the long magnetic lifetime of their electron quantum states. Accessing these states through electrical transport requires the engineering of their electronic configuration down to the level of individual atomic orbitals. Here, we address the magnetism of surface-supported lanthanide atoms, clusters, and films with orbital selectivity using X-ray absorption spectroscopy and magnetic circular dichroism. We exploit the selection rules of electric dipole transitions to reveal the occupation and magnetism of the valence electrons of Gd and Ho deposited on MgO/Ag(100). Comparing our results with multiplet calculations and density functional theory, we identify a charge transfer mechanism that leaves the lanthanide species in an unconventional singly ionized configuration. Our approach allows determining the role of valence electrons on the quantum level structure of lanthanide-based nanostructures.

## MAIN TEXT

### Introduction

In lanthanide elements, the spatial localization of electrons in the inner *4f* orbitals enables strong electron correlations and gives rise to large free-atom like spin and orbital magnetic moments. When surrounded by a suitable ligand environment, lanthanide atoms can exhibit long magnetic lifetime and offer an exquisite platform to explore information storage at the atomic scale *(1-6)*. Individual access and manipulation of their magnetic states have been shown using spin transport through electro-migrated molecular junctions *(7)* and spin-polarized scanning tunneling microscopy *(8-10)*. However, the strong localization of *4f* electrons severely limits their direct contribution to electron transport; hence, access to their magnetic states is readily available through their magnetically coupled valence electrons, either localized on the molecular ligand *(7, 11)* or on the outer orbitals of the atoms *(12, 13)*. Therefore, assessing the electronic configuration and spin polarization of the valence orbitals not only allows one to design molecular magnets with potentially accessible magnetic states *(14)* but also to rationalize the quantum level structure of surface adsorbed atoms with open valence shells *(9, 10)*. As for the latter, the case of Ho atoms on MgO has recently attracted particular attention due to the remarkable stability of their magnetic states up to 45 K *(3,8-10,15)*. Several works attempted to establish the ground state of these atoms, leading to two possible scenarios distinguished by different occupations of the *5d* states *(9)*. Recent observation of quantum tunneling of the magnetization in Ho atoms pointed towards an electronic configuration with no unpaired *5d* electrons *(10)*. However, a direct measurement of these electron states is still missing.

Here, we combine X-ray absorption spectroscopy (XAS) and magnetic circular dichroism (XMCD) with multiplet calculations and density functional theory (DFT) to address the magnetism of valence electrons in rare earth atoms, clusters, and films adsorbed on MgO/Ag(100). The optical selection rules of the electric dipole transitions allow the detection of the electron spin and orbital magnetic moments with orbital sensitivity *(16, 17)*. Conventional XAS/XMCD measurements of lanthanides generally make use of the *3d* → *4f* transitions to measure the large magnetic moment of the *4f* electrons. In this work, using an extended X-ray energy range, we were able to additionally access the higher energy *3p* → *5d* and *3s* → *6p* transitions to reveal the presence or absence of unpaired electrons in the valence *5d* and *6p* orbitals, respectively. To our knowledge, these measurements are unprecedented for surface-adsorbed single atoms and nanostructures. We first chose Gd as a reference system to test our combined experimental and theoretical approach due to its stable half-filled $4f^7$ configuration, weak magnetic anisotropy, and largest exchange splitting of the *5d* states *(18)*. We then applied our method to probe the valence electrons of Ho single atoms and determine their magnetic level scheme. For both elements, we reveal an unconventional singly ionized configuration, with bulk-like occupation of the *4f* shell ($4f^7$ for Gd, $4f^{10}$ for Ho), doubly occupied *6s* orbitals, and no unpaired electrons in the *5d* states for both systems. This latter evidence supports one of the proposed scenarios for the quantum levels of Ho atoms *(9,10)*, and clarifies the ground state electron configuration that determines their magnetic stability.

**Results**
The XAS and XMCD spectra were acquired at the BOREAS beamline of the ALBA synchrotron *(19)*, using the Hector end-station. As sketched in Fig. **1A,** we used left (LP) and right (RP) circularly polarized light in an external magnetic field B of up to 6 T parallel to the photon beam, while keeping the samples at 6.5 K. Spectra were acquired by sweeping the photon energy and collecting the signal in total electron yield mode. In this mode, the photo-ionized atoms at the surface are neutralized by an external electrical current, whose intensity is proportional to the photon absorption. A schematic of the X-ray absorption transitions discussed in this work is presented in Fig. **1B**. Optical selection rules for electric dipole transitions impose a change of the electron angular momentum $\Delta l = 1$ upon photon absorption. Hence, the choice of the core level determines the final state that can be probed using these transitions. Due to their larger absorption cross section, soft X-rays photons below 2000 eV are generally employed to probe diluted amounts of surface spins *(20-22)*. In the lanthanide series, this energy range allows reaching core levels with principal quantum number $n = 3$. Therefore, only *3s* → *6p* ($M_1$), *3p* → *5d* ($M_{2,3}$), and *3d* → *4f* ($M_{4,5}$) are the allowed electric dipole transitions that can be accessed using soft X-rays, with the double subscript index labeling the spin-orbit split core states, see Fig. **1B**. Although $M_{4,5}$ edges of rare earths are routinely measured in many soft X-rays beamlines, the larger energy needed to access the core levels at the $M_1$ and $M_{2,3}$ edges requires a specific design of the X-ray optics *(19)*. Here we measure the whole series of M transitions and thereby access the polarizations of both the inner and the outer shells.

*XAS and XMCD of Gd atoms and clusters*
Figure **2** shows an overview of these measurements for Gd atoms and Gd clusters adsorbed on MgO/Ag(100) (see Methods for the sample preparation, and section S1 for background

subtraction). In this figure, we report the sum (RP + LP) of the absorption signals from the two circular polarizations as XAS and their difference (RP-LP) as XMCD. Both $M_5$ and $M_4$ peaks (Figs. **2A-B**) consist of a fine multiplet structure with several narrow peaks whose intensity is determined by the interactions between the open *4f* and *3d* shells at the excited state *(23)*. As shown in Fig. **2B**, Gd atoms exhibit a large XMCD signal on both edges, with opposite sign between $M_5$ and $M_4$. This feature is a signature of the negligible orbital magnetic moment in this element, as expected for the case of half-filled *4f* shells *(16, 17)*. In order to ascertain the orbital configuration of Gd on MgO, we simulated the X-ray spectra using multiplet calculations including electron-electron interactions, spin-orbit coupling, crystal field resulting from the adsorption configuration, and external magnetic field (see Methods). The spectral features of both XAS and XMCD are well reproduced by multiplet simulations using the *$4f^7$* configuration. Angular and field dependent measurements of Gd *4f* electrons show negligible magnetic anisotropy, which is expected for such a half-filled shell configuration (see section S2 for angle-dependent measurement of atoms and clusters).

As visible in Figs. **2A** and **2C**, the intensity of the XAS signal decreases from the $M_5$ to the $M_1$ edge. This stems from the trend of decreasing number of core electrons from *3d* to *3s*, as well as the trend of decreasing electron orbital angular momentum involved in *3s → 6p*, *3p → 5d*, and *3d → 4f* transitions *(16, 17)*. In contrast to the $M_{4,5}$ edge, the XAS lineshapes of $M_{1,2,3}$ consist of a single broad peak. The absence of fine structure originates from the shorter lifetime of the final state that broadens the multiplet transitions into a single peak *(24)*. At the $M_3$, we observe a strong XMCD signal with the shape of a single peak derivative (Fig. **2D**). A slight dip is visible at the $M_2$ edge, but with reduced intensity with respect to $M_3$. Note that apart from the signal-to-noise ratio, the XAS and XMCD line shapes do not vary significantly from atoms to clusters and are well reproduced by our calculations using the *$5d^0$* configuration. As for the $M_1$, the extremely low signal allows the detection of this edge only for Gd clusters. The corresponding XMCD signal does not show any clear feature above our noise level, suggesting a negligible magnetic moment localized at the *6p* states. Comparison with simulated spectra from multiplet calculations assuming a *$6p^0$* configuration confirms this picture (Fig. **2D**).

*Understanding XAS and XMCD lineshapes of Gd $M_{2,3}$ edges*
To the best of our knowledge, XAS and XMCD at the $M_{2,3}$ edges of surface supported lanthanides have not been reported so far, hence, a specific analysis is required to unravel the underlying absorption mechanism. First, we discuss the lineshape of the dichroism as it is crucial to understand the magnetism and occupancy of *5d* shells. As a starting point, we interpret the peak derivative shapes observed in the XMCD of the $M_{2,3}$ edges based on a single-electron picture. Within this approximation, the circular polarization of the absorbed photons determines the spin-polarization of the excited photo-electrons *(17)*. According to the convention of magnetic field and photon axis shown in Fig. **1A**, LP photons at the $M_3$ edge excite a larger amount of spin-up electrons, while RP photons preferentially promote spin-down electrons *(25)*. As sketched in Fig. **3A**, the dichroism at this edge is hence expected to map the difference between the spin-up and spin-down *5d* density of states. Due to the spin splitting of the *5d* bands resulting from the intra-atomic interactions and orbital hybridization, the photon absorption produces two broad peaks centered at different energies for RP and LP. Hence the difference of the polarized XAS signal is expected to generate a peak derivative signal *(24)*, which we clearly observe at the $M_3$ edge of Gd/MgO, as illustrated in Figs. **3B-D**. Due to the reversed spin polarization of the photoelectrons at the $M_2$ edge, the XMCD signal (Figs. **3C-E**) is partially reversed with respect to the $M_3$. We observe these characteristic signals for Gd adatoms, clusters, and films on MgO (see section S3), as well as in reference samples of gadolinium(III) sulfate octahydrate (see section S4).

Even if the single-electron picture is sufficient for a qualitative interpretation of the XMCD lineshapes, a more detailed description including multi-electron configurations is required to ascertain the absence or presence of unpaired electrons in the *5d* shell. In Figs. **3D** and **3E,** we compare the results of the expected XMCD spectra for atomic $5d^0$ and $5d^1$ initial state configurations with the experiment. In order to compare the amplitudes, the sum of the two polarizations is matched to the total experimental XAS. As shown in Fig **3D**, the calculated XMCD at the $M_3$ edge shows an almost symmetric peak derivative shape for both $5d^0$ and $5d^1$ configurations. However, only the calculated spectrum of the $5d^0$ configuration accurately matches the amplitude of the experimental XMCD signal. The inclusion of an unpaired electron impacts the simulated XMCD spectrum at the $M_2$ edge even more prominently. As depicted in Fig **3E**, the XMCD lineshape would change from being an asymmetric peak derivative expected for an initial state configuration of $5d^0$ to a positive peak-like structure for $5d^1$. Such a peak at the $M_2$ edge would appear due to the large orbital angular momentum and charge asymmetry expected for an unpaired electron in the localized atomic orbital. Given the absence of such a signal in our measurements, we infer the absence of unpaired electrons in the *5d* shell of Gd atoms.

Our results suggest that the electronic structure of Gd atoms on MgO/Ag(100) is altered with respect to the free-atom ($4f^7 5d^1 6s^2 6p^0$) and to the bulk case *(26)*, both showing a net spin-polarization of the *5d* shells. The lack of *5d* polarization in surface-supported Gd atoms points towards a transfer of charge to the substrate. To validate this conjecture, we compare the XMCD of Gd atoms with that of clusters and a thin film, for which we expect to partially recover the *5d* polarization of the bulk. In order to capture the subtle variations occurring in the XMCD spectra, we perform an analysis of the integrals of XMCD edges, which are related to the spin and orbital magnetic moments of the target orbitals projected along the photon-beam axis *(16, 17)*. As shown in Fig. **3F** the area of the XMCD at the $M_3$ edge is positive for the atoms and decreases to a negative value when increasing the Gd coverage. Conversely, the area of the XMCD at the $M_2$ edge remains negative for all Gd coverages (Fig. **3G**). According to the sum rules, the absence of unpaired electrons in the ground state should naturally lead to vanishing integrals of the XMCD peaks, owing to the absence of net spin and angular momentum. Otherwise, parallel alignment of the spins to the outer magnetic field normally results in a negative value of the $M_3$ integral *(16, 17)*. Hence, the trend to negative integral values at the $M_3$ edge indicates an increase of the spin magnetic moment and occupancy of the *5d* states with increasing Gd coverage. On the other hand, the positive integral at the $M_3$ edge of Gd does not have a straightforward interpretation in terms of sum rules analysis. Remarkably, our multiplet calculations predict a positive value of the $M_3$ XMCD integral, although not fully matching the experimental value. By selectively suppressing the interactions between the open shells in our calculations, we identify the origin of this effect in an indirect *4f-3p-5d* coupling, which partially mixes the polarization of the *4f* and *5d* electrons in the final state. The same effect partially explains the negative value of the integral at the $M_2$ edge. Additional contribution to the XMCD integral can be ascribed to a so-called breathing effect, namely an exchange-driven contraction of the valence *d* shells *(24)*. As shown in previous measurements at the $L_{2,3}$ edges (*2p* → *5d* transitions) of Gd bulk metal *(27, 28)* and compounds *(25, 29-31)*, this effect can introduce a spin-dependence in the matrix element of the X-ray transitions, which produces an asymmetry in the derivative peak shape and, consequently, a net component of the XMCD integral. Due to these effects, a simple sum rules analysis cannot be performed on the $M_{2,3}$ edges of Gd. Nevertheless, the possibility to quantify trends in the XMCD integrals extends the applicability of our approach to a wide variety of systems, from atoms with localized electrons to metal films with dispersive bands.

*Density functional theory of Gd atoms*
In order to identify the origin of the charge transfer mechanism discussed in the previous section and to understand the individual roles of the MgO film and supporting Ag(100) substrate, we performed density functional theory (DFT) calculations of the Gd atoms adsorbed on the oxygen-top site of both bare MgO and MgO/Ag(100) *(3)*. We found that Gd adsorbs approximately 2.71 Å above the top layer of MgO and the underlying oxygen is displaced towards the Gd resulting in a Gd-O bond length of 2.15 Å, as illustrated in Fig. **4E**. To gain insight in the resulting electronic structure, we calculated the projected density of states (PDOS) for both Gd/MgO (Fig. **4A**) and Gd/MgO/Ag(100) (Fig. **4B**). For both cases, the related Löwdin charge analysis (Table **1**) finds seven electrons in the *4f* shell, in agreement with the experiment. The PDOS plots reveal full polarization of these states, with the majority *4f* electrons giving rise to a large localized magnetic moment. In addition, both of these calculations reveal equally occupied *6s* majority and minority orbitals lying below the Fermi level, with resulting negligible net polarization (Fig. **4A** and **4B**). The *6p* orbitals are essentially empty and positioned at about 2 eV above the Fermi level.

On the other hand, the total Gd charge as well as the occupation of the *5d* orbitals is strongly affected by the Ag(100) substrate. For the bare Gd on MgO system, we find the total charge state of Gd on MgO to be identical to the free neutral atoms. Differently from the free atom, we find an electronic reconfiguration occurring mostly between the *6s* and *5d*, which results from the on-site hybridization upon adsorption at the MgO surface. This electron redistribution, however, does not alter the total polarization of the outer shells, leaving essentially one unpaired electron in the *5d* shell as sketched in panel **4C**. Conversely, when the silver substrate is included, we find an effective charge transfer of 0.80 electrons to the substrate, which essentially coincides with the variations occurring in the Gd *5d* orbital (-0.83 electrons). The main variation in the electron states occurs in the planar $5d_{x^2-y^2}$ orbital, which is the lowest in energy in the MgO crystal field (see section S5). Due to this charge rearrangement, the *5d* states become essentially non-polarized (Figs. **4B** and **4D**), in agreement with our measurements. The charge is mostly transferred to the states localized at the MgO/Ag(100) interface (Fig. **4E**). Such an effect has been quite commonly observed to occur in ultrathin oxide layers on metal surfaces *(32)* as well as on metal-supported graphene layers *(13)*. Due to this charge transfer, the atom is left in a formal "Gd$^{+1}$" charge state.

Differently from the *$5d^0$* configuration inferred from the multiplet analysis, DFT finds a residual charge of 0.7 electrons in the *5d* orbitals even in the presence of Ag(100). This value is mostly a result of the hybridization between Gd and MgO states. The discrepancy between the two analyses originates from the absence of covalence in the multiplet approach that prevents an accurate modelling of the ligand charges. However, since this fraction of *5d* electrons has only negligible spin-polarization, it does not provide a significant contribution to the XMCD signal. In addition, no strongly polarized state is localized at the Fermi level, with the lowest majority *5d* states available at around +200 meV. The absence of polarized valence electrons is expected to result in a small magnetic signal in transport measurements.

*Ho atoms and clusters*
We now apply the understanding gathered from the orbital-sensitive M-edges measurements of Gd atoms and clusters on MgO to explore the orbital occupation and magnetism of Ho single atom magnets. The normalized XAS and XMCD at the $M_{4,5}$ edges (see section S6) indicate *$4f^{10}$* occupancy for Ho atoms, in agreement with earlier work *(3)*. The measurements of the $M_{2,3}$ edge show similar behavior to what we have observed for Gd, *i.e.*, derivative-like XMCD at the $M_3$ edge and an absence of a peak structure at the $M_2$ edge XMCD (Fig. **5A-B**). The magnitude of the XMCD at both edges is, however, quite reduced with respect to Gd. This effect is due to the smaller number of unpaired electrons at the *4f* orbitals that induce a smaller exchange splitting of the *5d* bands *(18)*. The reduced intensity of the XMCD at these edges is well matched by our multiplet calculations. For the case of Ho atoms, the signal-to-noise ratio at the $M_3$ edge does not allow us to unambiguously distinguish the presence or absence of an unpaired electron in the *5d* states. On the other hand, the comparison with *$5d^1$* calculations is particularly mismatched at the $M_2$ edge, where we only find good agreement when no unpaired electrons are present in the *5d* orbitals.

The similarities with Gd suggest a *$6s^2$* and *$6p^0$* configuration also for the Ho atoms, with no unpaired electrons in the outer shells, as summarized in Fig **5C**. In this electronic configuration, multiplet calculations predict a ground state multiplet with maximum spin **S** = 2, orbital **L** = 6, and total angular momentum **J** = 8, as also expected by Hund's rules. The crystal field provided by the MgO splits the lowest 17 states of this multiplet, as shown in Fig. **5D**. In absence of singly occupied *5d* electrons, the main contribution to the splitting comes from the proximity of the oxygen ion underneath the Ho, which generates a large uniaxial anisotropy and favors a ground state doublet with maximum projection $\langle J_z \rangle \approx \pm 8$ along the out-of-plane direction. For such a ground state, we find a total magnetic moment of 9.9 $\mu_B$, which agrees with previous STM results *(8, 9)*. The perturbation of the neighboring Mg and O ions enables quantum tunneling of the magnetization and further splits originally conjugated doublets into pairs of singlet states with $\langle J_z \rangle \approx \pm 0$ at the top of the barrier. As discussed in previous works, these states can offer effective pathways to reverse the magnetization of the atom using high-energy electrons *(8, 9)*. Nevertheless, our calculations show that the ground state doublet preserves a very strong axial character, *i.e.* its value of $\langle J_z \rangle$ is very close to the maximum theoretical value and quantum tunneling of magnetization is very weak for these states.

As discussed for Gd, the absence of polarized *5d* electrons is expected to result in a very small magnetic contribution to the conductance in transport measurements. Previous spin-polarized STM results on Ho atoms on MgO/Ag(100) found a rather low magnetic contrast of about 4% of the total conductance *(8 - 10)*. This value is an order of magnitude smaller than for Fe on the same substrate *(33)* and for Dy atoms on graphene/Ir(111), for which an unpaired electron in the *6s* orbitals provides a large magnetic contribution to the tunneling conductance *(13)*.

**Discussion**

The combination of XMCD measurements with density functional theory and crystal-field based multiplet analysis allows us to characterize the magnetism of surface adsorbed atoms and clusters with orbital sensitivity. For both Gd and Ho, our observations corroborate a bulk-like *4f$^n$* configuration (*n* = 7 for Gd, *n* = 10 for Ho). For neutral species, that would require a singly occupied *5d* orbital. However, we found no evidence of unpaired electrons in the valence shell of these late lanthanides. The absence of polarization of the *5d* states is ascribed to an effective charge transfer of almost one electron from the rare earth adatoms to the Ag(100) substrate.

The single ionized state observed here is quite unconventional in rare earth elements, which are mostly found in neutral, divalent or trivalent states *(34)*. The key to obtain this unconventional configuration is the presence of an ionic support that offers a single uniaxial bond to the adatom, with limited possibility of orbital hybridization and charge transfer. In this situation charge transfer can only occur due to the proximity of the underlying metal substrate. Consequently the rare earth electron tunnels through the MgO film and reaches the empty states of Ag(100). The resulting low polarization of the outer shells explains previous STM results showing low magnetic contrast on lanthanide atoms on MgO/Ag(100) *(8 - 10)*.

The absence of unpaired electrons in the *5d* shell of Ho atoms eliminates one of the two quantum level configurations proposed in a previous report *(9)*. In the *4f$^{10}$6s$^2$* configuration inferred from this work, the effect of the MgO crystal field acting on the *4f* orbitals favors the configuration with the largest projection of the total angular momentum $\langle J_z \rangle \approx \pm 8$. Although such a ground state is not considered magnetically stable in the four-fold symmetric ligand field of the oxygen-top adsorption site, the strong uniaxial component of the crystal field dominates over the transverse perturbative terms. In this configuration, a limited but detectable quantum tunneling can limit the stability of the system at very low fields, as also found in a previous work *(10)*.

In conclusion, the possibility of investigating surface-supported nanostructures with orbital selectivity allows to address charge transfer and electronic reconfiguration effects that crucially impact the structure of quantum levels. Our study outlines a strategy to design lanthanide-based atomic-scale systems with highly localized states utilizing exchange-coupled valence orbitals as a read-out channel in electrical transport measurements.

**Materials and methods**

Single crystals of Ag(100) were prepared by repeated cycles of sputtering and subsequent annealing at 773 K. Films of MgO with thickness between 3 and 6 monolayers (MLs) were grown by thermal evaporation of Mg in $O_2$ partial pressure of $1\times10^{-6}$ mbar, with the substrate kept at 623 K and a Mg flux yielding a growth rate of about 0.2 ML/min. One monolayer is defined as one MgO(100) unit cell per Ag(100) substrate atom. The calibration of the MgO thickness was obtained by measuring the Mg XAS K edge and comparing to previous works *(3, 15)*. The MgO/Ag(100) samples were transferred to the measurement position without breaking the vacuum. Gadolinium and holmium atoms were deposited from thoroughly degassed lumps (purity 99 %) inserted into W crucibles. The lanthanides elements were deposited on the substrate held at less than 10 K and in a base pressure of $1\times10^{-10}$ mbar. Also for the adsorbed elements, we define 1 ML as one lanthanide atom per Ag(100) substrate atom, and calibrated the amount by comparison with former experiments *(3, 15)*.

The XAS and XMCD spectra shown in the main text were acquired at the BOREAS beamline of the ALBA synchrotron *(19)*, with additional measurements done at the EPFL/PSI X-Treme beamline at the Swiss Light Source *(35)* and at the 6A MeXiM beamline at Pohang Light Source-II. The XAS/XMCD acquisitions were performed with circularly polarized light in the total electron yield (TEY) mode at sample temperatures of T = 6.5 K (BOREAS),  T = 2.5 K (X-Treme) or T = 15 K (MeXiM), and in external magnetic fields up to B = 6.8 T parallel to the x-ray beam. Two different incidence geometries have been used to explore the angular dependence of the magnetic moments (see section S2), namely normal incidence, with both beam and field perpendicular to the surface, and grazing incidence, with the sample rotated by 60 degrees with respect to the beam and field direction.

In order to isolate the contribution of the lanthanide atoms and clusters from the background signal, spectra of bare MgO/Ag(100) over the lanthanide edges were recorded prior to Gd and Ho deposition and subtracted from the final spectra (see section S1). The spectra at different edges of the atoms have been normalized to the value of the integrated XAS at the $M_{4,5}$ edges. In order to compare the $M_{2,3}$ edges of atoms, clusters, and films, the values have been normalized to the summed XAS over these two edges.

DFT calculations were performed using plane-wave basis and pseudopotentials as implemented in Quantum Espresso with pseudopotentials from the SSSP library *(36, 37)*. All pseudopotentials use the generalized gradient approximation (GGA) for the exchange-correlation potential. The sample was modeled as 4 monolayers (ML) of silver fixed at the lattice constant of bulk silver ($a_{Ag}$=4.16 Å) capped by 3 ML of MgO and the adatom which is laterally separated from its period images by more than 12 Å. The cell is padded by 15 Å of vacuum to decouple the slab from its periodic image in z-direction. Integration of the Brillouin zone used a 5x5x1 Monkhorst-Pack grid *(38)* and a cutoff of 50 Ry and 500 Ry was used for the plane wave expansion and the charge density, respectively. We performed a relaxation of the system until the residual forces were less than $10^{-3}$ Ry/$a_0$ ($a_0$ is the Bohr radius). To corroborate our finding of charge transfer from the ad-atom to the underlying Ag substrate we performed additional calculations after removing the silver. This can be interpreted as the limiting case of very thick MgO films. The structure was not further relaxed and all other parameters are identical to the calculations that include the silver layer.

Simulated X-ray spectra were computed using the Quanty multielectron code *(39)*. Open shell multiplet calculations of the Gd and Ho atoms were performed by modeling the intra- and inter orbital interactions using Slater integrals, whose values were obtained from full-electron Cowan's atomic structure code *(40)*. For both Gd and Ho, all Slater integrals were calculated assuming doubly occupied *6s* orbitals, *4f$^n$* configuration with n = 7 for Gd and n = 10 for Ho and empty *6p* shells. The values were rescaled to account for the screening provided by surface electrons. Rescaled values are provided in Table S2. Simulations of both empty and singly occupied *5d* orbitals were performed as indicated in the text. In addition to electron-electron interactions, the Hamiltonian for the multiplet calculations also includes spin-orbit coupling, Zeeman energy due to the external magnetic field and crystal field acting on the different shells. Spin orbit values were computed using Cowan's atomic structure code. The final state Hamiltonian includes the presence of the core hole and additional valence electron according to the simulated transitions.

The crystal field acting on the open shells of lanthanide was modeled according to the nature of the open valence shell. For delocalized *5d* and *6p* electrons, the splitting of the orbitals was obtained from DFT by means of Wannier functions method (see section S5). This method is however not accurate enough to be applied to the highly localized and correlated *4f* electrons. In this case, we modeled the crystal field acting on the *4f* orbitals using a point charge electrostatic model *(41, 42)* including only the nearest neighbors of the adsorbed lanthanide. The position and charge values of the neighboring ions around the Gd and Ho atoms obtained from DFT are provided in Table S3. The value of the crystal field parameters used for the calculations are summarized in Table S4, S5, and S6. The XAS and XMCD spectra were calculated using the Green's-function method *(39)*.

**Acknowledgements**

**Funding:** A.S., D.S., C.W., S.L.A., D.K., L.C., A.J.H., and F.D. acknowledges funding from the Institute of Basic Science, Korea through the Project No. IBS319 R027-D1. W.S.N. and J.H.P. acknowledges funding from the National Research Foundation of Korea (NRF) funded by the Ministry of Science and ICT (No. 2016K1A4A4A01922028).

**Author contributions:** A.B. and F.D. conceived the experiment. A.S., S.L.A., D.K., P.G., A.B., and F.D. performed the experiments at the BOREAS beamline of the ALBA synchrotron. A.S., S.L.A., D.K., M. P., S.R., A.B., and F.D. performed the experiments at the EPFL/PSI X-Treme beamline of the Swiss Light Source. D.S, C.W., S.L.A., D.K., L.C., W.S.N., and F.D. performed the experiments at 6A MeXiM beamline at Pohang Light Source-II. A.S., D.S., S.L.A., and F.D. analyzed the data. A.S., D.S., S.A. and F.D. performed the multiplet calculations. C.W. performed and analyzed the density functional theory calculations. J.H.P., H.B., A.J.H. secured funding and supervised the project. All authors discussed the results and analysis. A.S., D.S., C.W., and F.D wrote the manuscript with the input of all other authors.

**Figures and Tables**

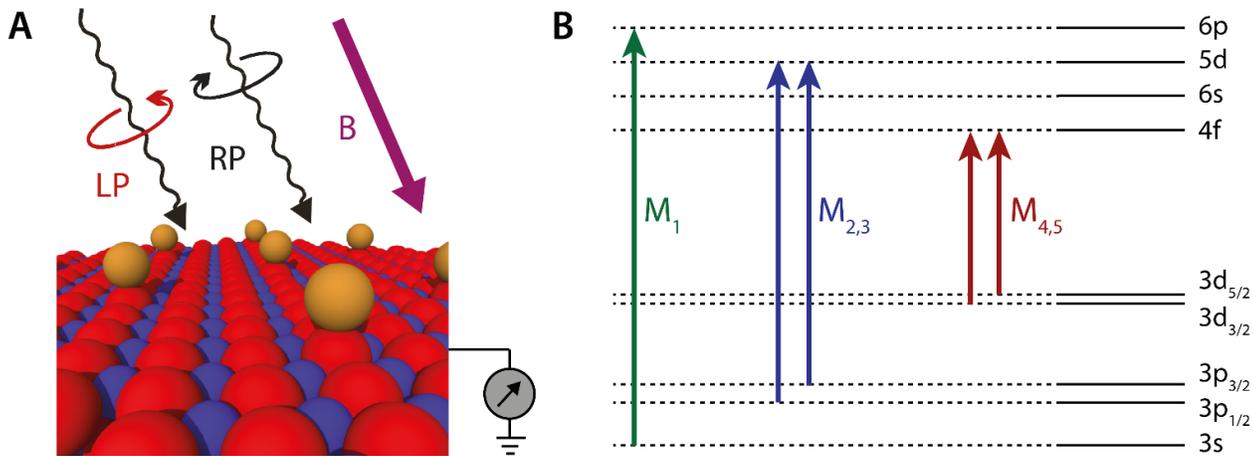

**Figure 1: Orbital-selective X-ray absorption on single atoms. (A)** Schematics of the XAS/XMCD measurements performed in total electron yield mode. **(B)** Representation of orbital sensitive measurements at $M_1$, $M_{2,3}$ and $M_{4,5}$ edges, which address the occupancy and magnetism of the *6p*, *5d*, and *4f* orbitals, respectively.

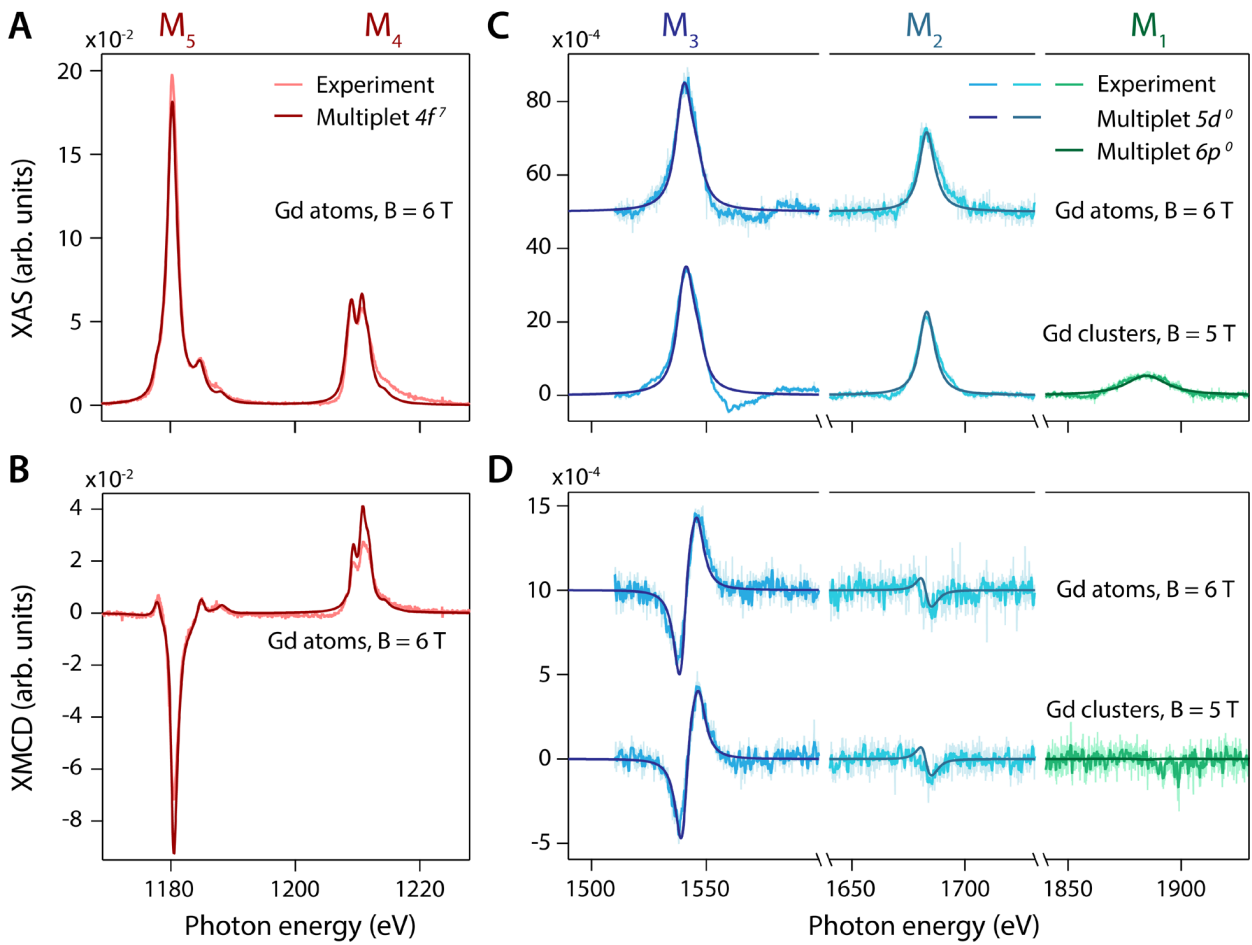

**Figure 2. X-ray absorption spectra of Gd on MgO/Ag(100). (A - B)** XAS (RP + LP) and XMCD (RP - LP) of Gd atoms (0.03 ML) at the $M_{4,5}$ edges probing the *4f* orbitals. **(C-D)** XAS and XMCD of Gd atoms (0.03 ML) and clusters (0.14 ML) at the $M_{2,3}$ and $M_1$ edges probing the *5d* and *6p* orbitals, respectively. Raw (light solid lines) and smoothened data (solid lines) are shown together with multiplet calculations for a *4f $^7$ 6s$^2$ 5d $^0$ 6p$^0$* configuration. Arbitrary units refer to the same intensity scale. All data were acquired in grazing incidence (see Methods).

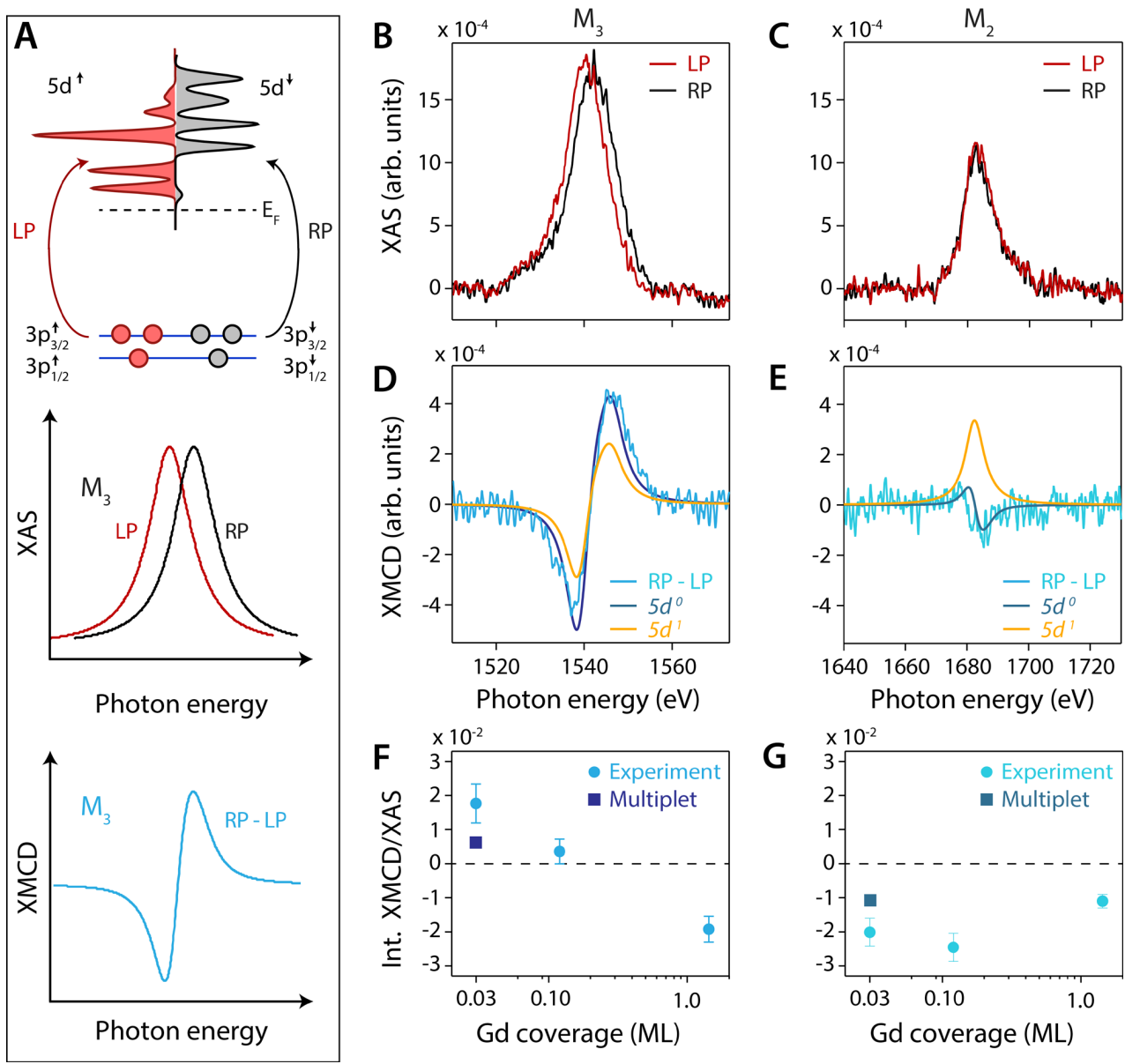

**Figure 3: Analysis of the $M_{2,3}$ edges of Gd atoms. (A)** Schematics of X-ray absorption at the $M_3$ edge. The preferential spin-polarization generated by RP and LP photons probes the spin-split $5d$ states of Gd obtained from DFT (see Fig. **4B**). The short lifetime of the excitation gives rise to broad peaks in the XAS. The difference of XAS signals from the two polarizations results in a derivative-like XMCD. **(B-C)** Smoothened XAS of Gd atoms on MgO/Ag(100) at the $M_3$ and $M_2$ edges. **(D-E)** Corresponding XMCD signals are compared to simulations for both $5d^0$ and $5d^1$ configurations. **(F-G)** Integrated XMCD/XAS signals for atoms (0.03 ML) clusters (0.14 ML) and films (1.2 ML) at the $M_3$ and $M_2$ edges. All data were acquired in grazing incidence (see Methods).

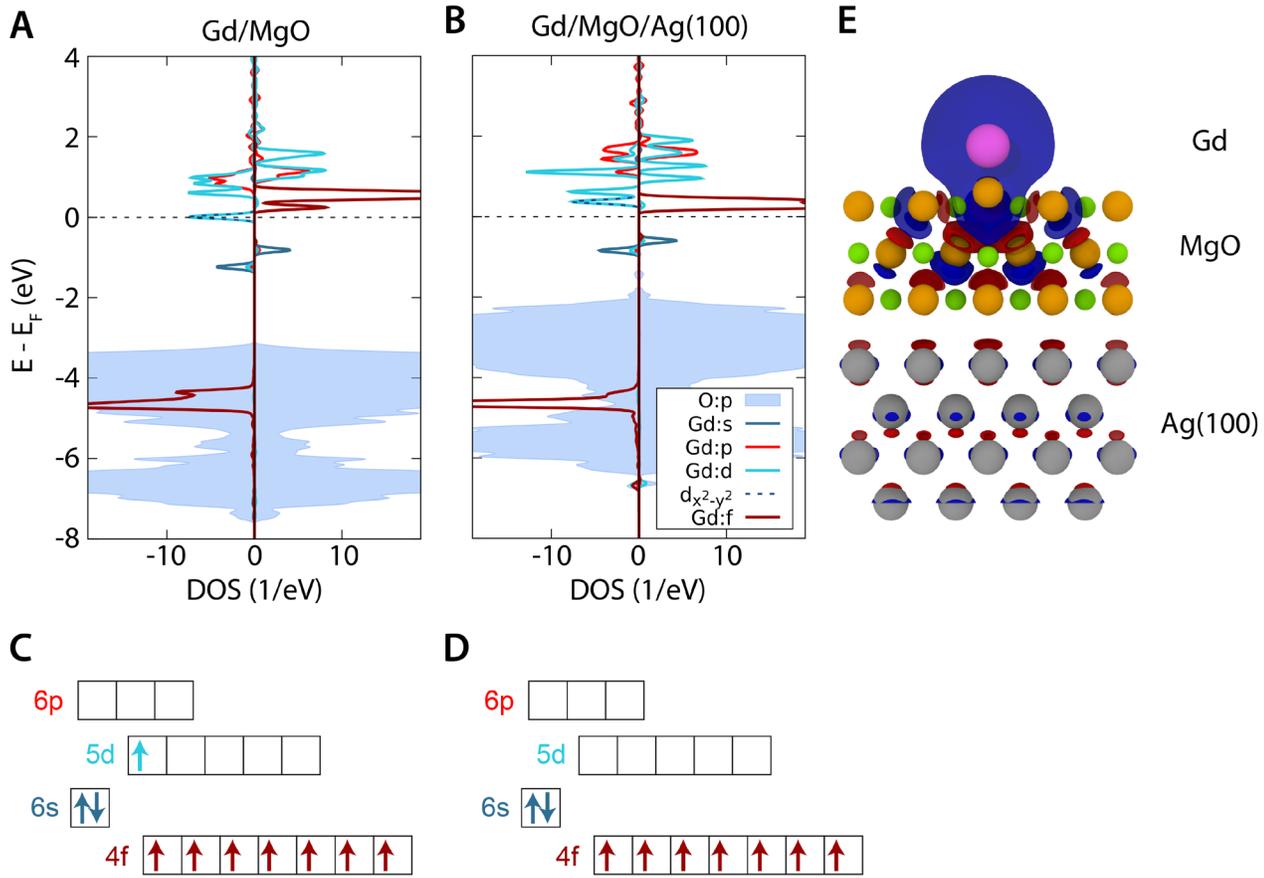

**Figure 4: Density functional theory of Gd atoms on MgO/Ag(100). (A)** Local spin-resolved DOS of the adatom on the oxygen-top site of 3 ML MgO, and **(B)** on the oxygen-top site of 3 ML MgO/Ag(100). **(C-D)** Schematics of the electronic configuration of Gd on MgO without and with the Ag(100) substrate. **(E)** Cross-section along the MgO [110] direction showing the Gd/MgO/Ag(100) slab. Overplotted is the isosurface of charge density difference obtained from subtracting the electron density of the individual Gd and MgO/Ag(100) to the whole Gd/MgO/Ag(100) slab. Red/blue regions indicate positive/negative electron charge variations at the isosurface of $6 \times 10^{-4}$ $e/a_0^3$ ($a_0$ being the Bohr radius). The electron charge transfer from the Gd adatom to silver can be seen in the red regions around the top silver layer atoms.

| Orbital | Free atom | Gd/MgO | Gd/MgO/Ag(100) | Charge difference |
|---|---|---|---|---|
| *6s* | 2.0   (0.0) | 1.06   (0.01) | 1.10   (0.02) | 0.04 |
| *6p* | 0.0   (0.0) | 0.24   (0.05) | 0.23   (0.00) | -0.01 |
| *5d* | 1.0   (1.0) | 1.54   (1.04) | 0.71   (0.08) | -0.83 |
| *4f* | 7.0   (7.0) | 7.16   (6.99) | 7.16   (6.93) | 0.00 |
| Total | 10.0   (8.0) | 10.00   (8.09) | 9.20   (7.03) | -0.80 |

**Table 1: Electron configuration in Gd atoms.** DFT calculated effective charges for Gd free-atom, adsorbed on bare MgO, and on MgO/Ag(100) calculated with the Löwdin scheme. The expected charge difference due to the presence of the Ag substrate is given in the last column. All values are provided in electron charges. For the *6s* and *6p* orbitals, the charges have been obtained by subtracting 2 and 6 electrons from the total charge obtained for the *5s+6s* and *5p+6p*, respectively, as obtained from the Löwdin analysis.

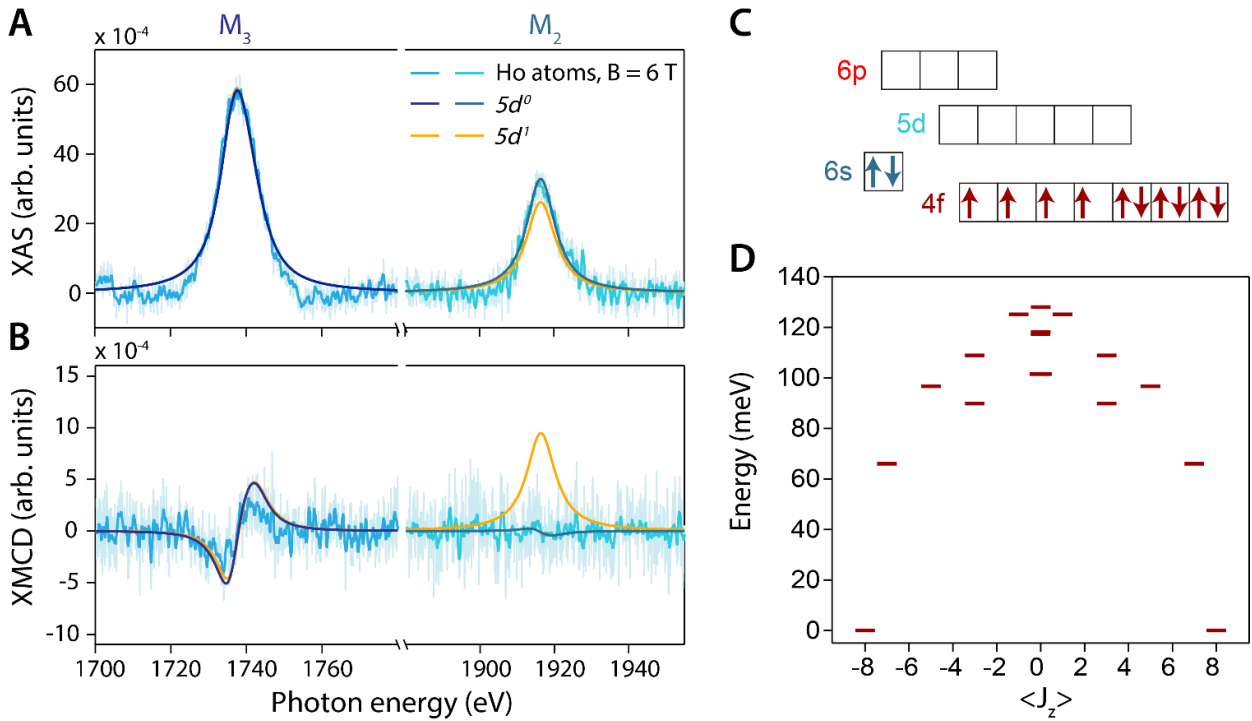

**Figure 5: XAS and XMCD of Ho atoms on MgO/Ag(100). (A-B)** XAS and XMCD spectra for Ho atoms (0.03 ML) at the $M_{2,3}$ edges. Raw (light solid lines) and smoothened data (solid lines) are shown together with multiplet calculations for $5d^0$ and $5d^1$ configurations. **(C)** Orbital occupation of Ho atoms inferred from the analysis of the $M_{2,3}$ edges. The lowest energy multiplet is characterized by S = 2, L = 6, J = 8 quantum numbers. (c) Splitting of the magnetic states from the lowest multiplet due to the crystal field, revealing a ground state with $J_z \approx \pm 8$. To maximize the magnetic signal in the presence of the Ho out-of-plane anisotropy *(3)*, all data were acquired in normal incidence (see Methods).

# Supplementary Information for "Orbital-resolved single atom magnetism measured with X-ray absorption spectroscopy"

Aparajita Singha[1,2,†], Daria Sostina[1,2,†], Christoph Wolf[1,2,†], Safa L. Ahmed[1,3], Denis Krylov[1,2], Luciano Colazzo[1,2], Pierluigi Gargiani[4], Stefano Agrestini[4], Woo-Suk Noh[5], Jae-Hoon Park[5,6], Marina Pivetta[7], Stefano Rusponi[7], Harald Brune[7], Andreas J. Heinrich[1,3], Alessandro Barla[8], and Fabio Donati[1,3,*]

[1]*Center for Quantum Nanoscience, Institute for Basic Science (IBS), Seoul 03760, Republic of Korea*
[2]*Ewha Womans University, Seoul 03760, Republic of Korea*
[3]*Department of Physics, Ewha Womans University, Seoul 03760, Republic of Korea*
[4]*ALBA Synchrotron Light Source, Cerdanyola del Vallès, 08290, Catalonia, Spain*
[5]*MPPC-CPM, Max Planck POSTECH/Korea Research Initiative, Pohang 37673, Korea*
[6]*Department of Physics, Pohang University of Science and Technology (POSTECH), Pohang 37673, Korea*
[7]*Institute of Physics, École Polytechnique Fédérale de Lausanne, Station 3, CH-1015 Lausanne, Switzerland*
[8]*Istituto di Struttura della Materia (ISM), Consiglio Nazionale delle Ricerche (CNR), I-34149 Trieste, Italy*

[†] *Equally contributing authors*
*E-mail: donati.fabio@qns.science

S1. Background subtraction in x-ray absorption spectra
S2. Angle-dependent measurements of Gd on MgO/Ag(100)
S3. XAS and XMCD of Gd film on MgO/Ag(100)
S4. Reference measurements on Gd sulfate octahydrate
S5. Details of DFT calculations
S6. Additional data for Ho atoms on MgO/Ag(100)
S7. Details of multiplet calculations
S8. Supplementary references

## S1. Background subtraction in x-ray absorption spectra

In order to compare the shape of the x-ray absorption (XAS) and magnetic circular dichroism (XMCD) spectra measured on different samples, we performed the following analysis. First, all spectra were subtracted from a respective background profile, which was acquired prior to the deposition of the rare earth atoms on the MgO substrate, without changing any other experimental parameters. The background subtracted absorption signals are shown in Figure **S1 A** and **C** for $M_3$ and $M_2$ absorption edges respectively, from 0.14 ML Gd on MgO/Ag(100). Note that the XAS of background and atoms were normalized such that they match at the two ends of the spectral range for the corresponding x-ray polarization. Second, the baseline of thus obtained absorption spectra exhibited a step-like feature as evident in Figure **S1 B**. Very similar behavior has been previously observed for $L_{2,3}$ edges of Gd *(30)*. This originates from non-resonant excitations, namely when the core electrons are excited with sufficient energies allowing access to the continuum of unoccupied states. Note that this effect is less evident for the $M_{4,5}$ edges as the signal from the corresponding absorption edge overwhelms the overall effective spectral shape.

For all edges, we subtracted this step profile by fitting the total absorption, defined by the sum of the signals from two circular polarizations, with the following function:

$$f(x) = a + b*x + c*x^2 + d*\left[1 - \frac{1}{1+\exp(\frac{x-e}{w})}\right].$$

Here $a$, $b$, $c$, $d$ and $e$ are fitting parameters, while $w$ is the step-width, which was chosen to be 7 eV for both Gd and Ho for $M_{1,2,3}$ absorption edges. This allowed us to obtain an almost featureless background for all energies except for the position of the respective absorption peak. The remaining post-edge oscillatory absorption profile comes from extended x-ray absorption fine structure (EXAFS). An example of subtracting such a step profile is shown in Figure **S1 B** and **D** for $M_3$ and $M_2$ edge data acquired from 0.14 ML Gd on MgO/Ag(100).

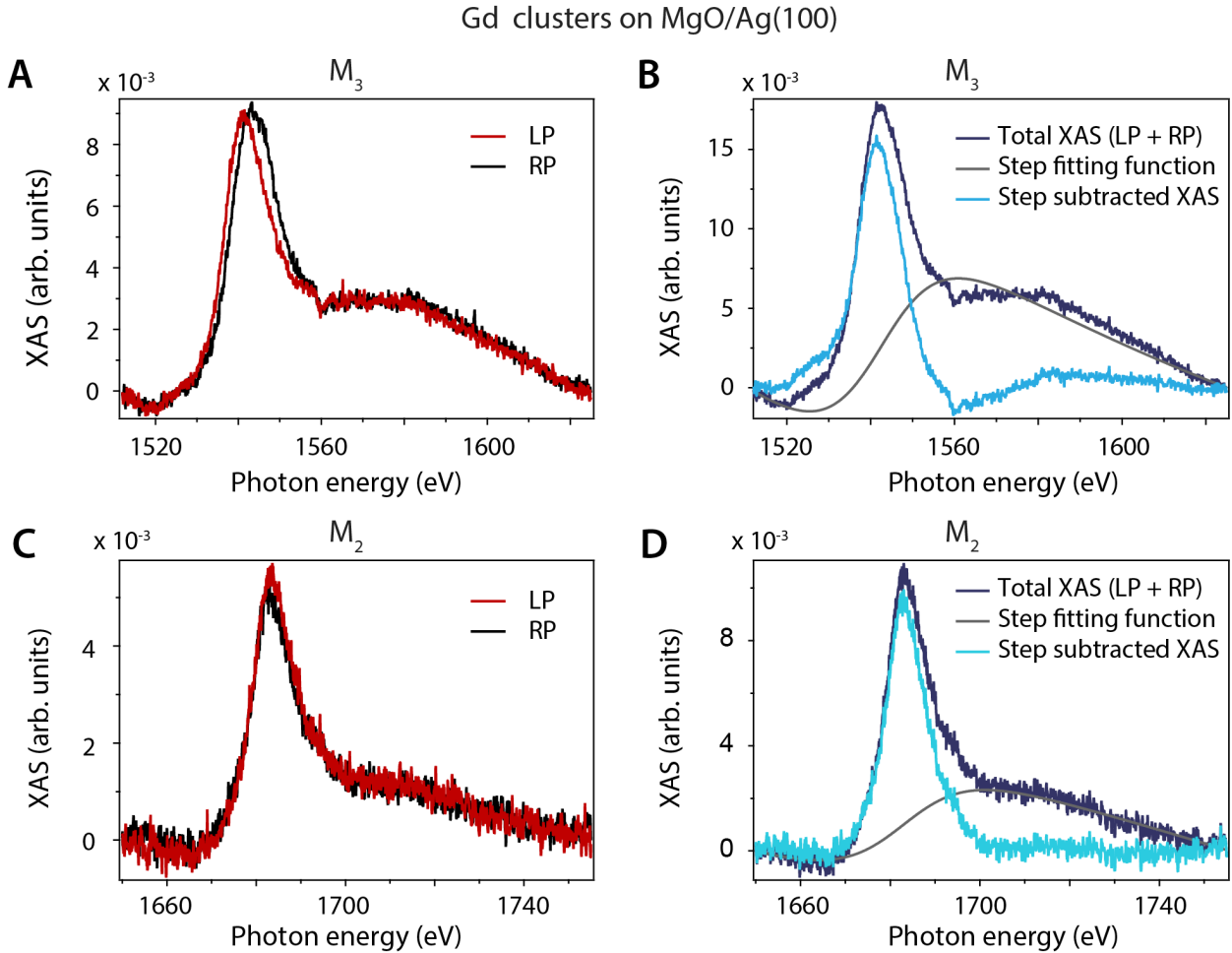

*Figure S1: Background subtraction at the $M_{2,3}$ edges. A: Left (LP) and right polarization (RP) XAS signal of the $M_3$ edge of Gd clusters on MgO/Ag(100) after subtraction of the bare MgO/Ag(100) background signal. B: The total XAS signal obtained by summing LP and RP from A is shown together, with a step fitting function. The resulting step subtracted XAS is also shown. C and D: Same procedure of A and B is reported for the $M_2$ edge. All measurements are performed in grazing incidence (B= 5 T, T = 6.5 K, Gd coverage $\Theta_{Gd}$ = 0.14 ML).*

Finally, in order to compare the spectral shape of XMCD from different atom and cluster samples we normalized all spectra with respect to integrated XAS at the $M_{4,5}$ edge measured on the same sample.

### S2. Angle-dependent measurements of Gd on MgO/Ag(100)

We performed further characterization of the magnetism of Gd atoms and clusters on MgO/Ag(100) by angle-dependent XAS and XMCD measurements. As shown in Fig. **S2A** and **B**, the Gd spectra at the $M_{4,5}$ edges at the maximum available field are essentially identical along the two indicated directions, namely normal incidence (photon beam and magnetic field normal to the sample surface) and grazing incidence (both photon beam and field at 60 degrees from surface normal). Magnetization loops along the two directions show slight differences, with a weak out-of-plane easy axis, see Fig. **S2 C**. The weak anisotropy is a general property of half-filled 4f shell, with the total charge close to spherical symmetry *(43)*.

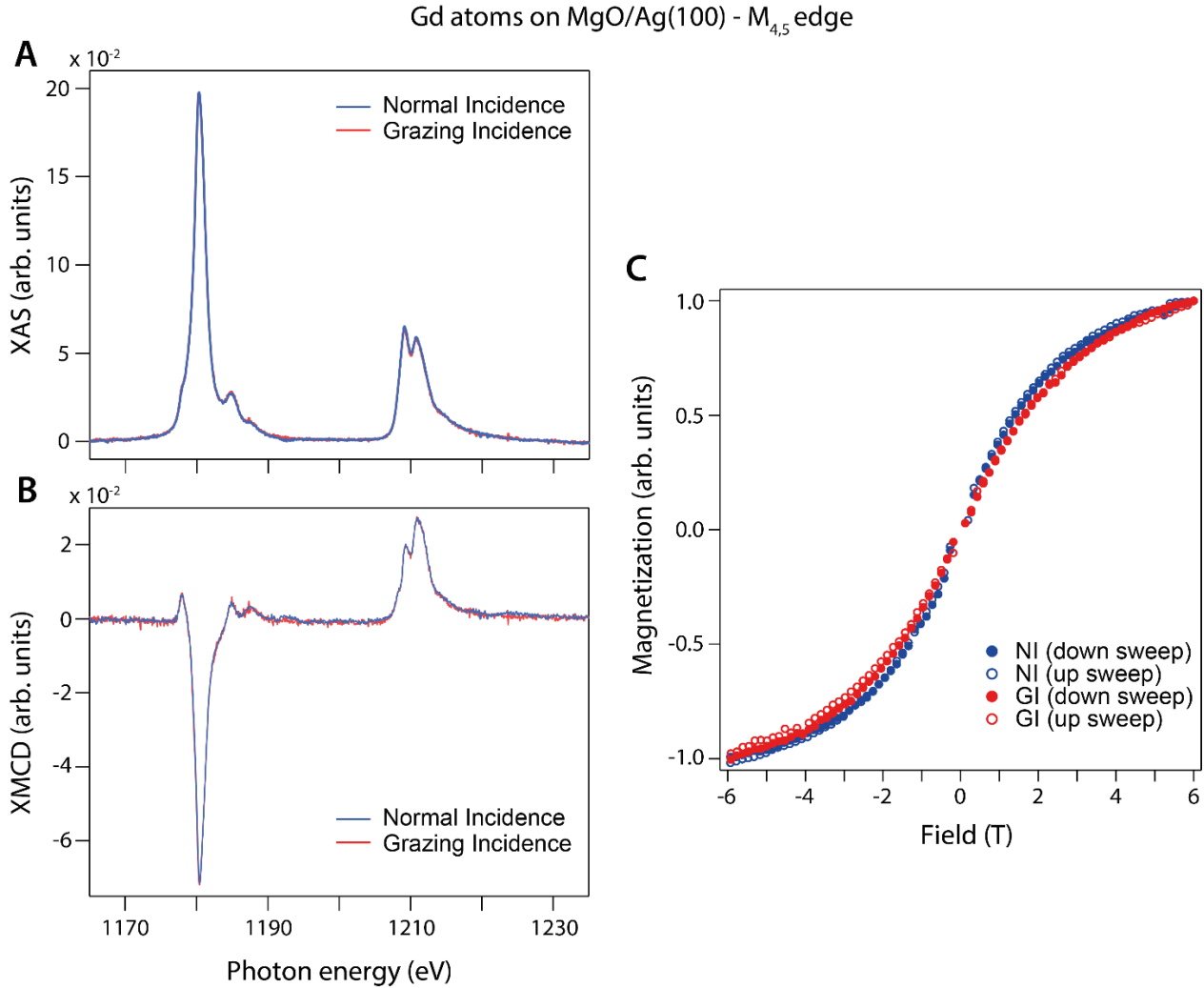

*Figure S2: Anisotropy of Gd atoms on MgO/Ag(100).* **A**: *XAS and* **B**: *XMCD measurements at the $M_{4,5}$ edge of Gd for two incident angles, normal incidence (NI) and grazing incidence (GI)) at B= 6 T.* **C**: *Magnetization loops of the Gd atoms measured at the $M_5$ edge. All measurement performed at T = 6.5 K ($\Theta_{Gd}$ = 0.03 ML).*

Gadolinium clusters also show negligible angular anisotropy in the XAS and XMCD spectra at B = 5 T, see Fig. **S3 A** and **B**. Due to the larger absorption signal available for this sample with larger Gd coverage, we compared the amplitude of the field dependent XMCD signal obtained at the $M_5$ and $M_3$ edges (Fig. **S3 C**). The two signals show the same field dependence, indicating strong magnetic coupling between the 4f and 5d states. We note that even in the absence of 5d electrons in the ground state, the XMCD signal is sensitive to the magnetic coupling at the excited state *(44)*, which in our case is represented by an additional photoelectron promoted from the 3p to the 5d orbitals.

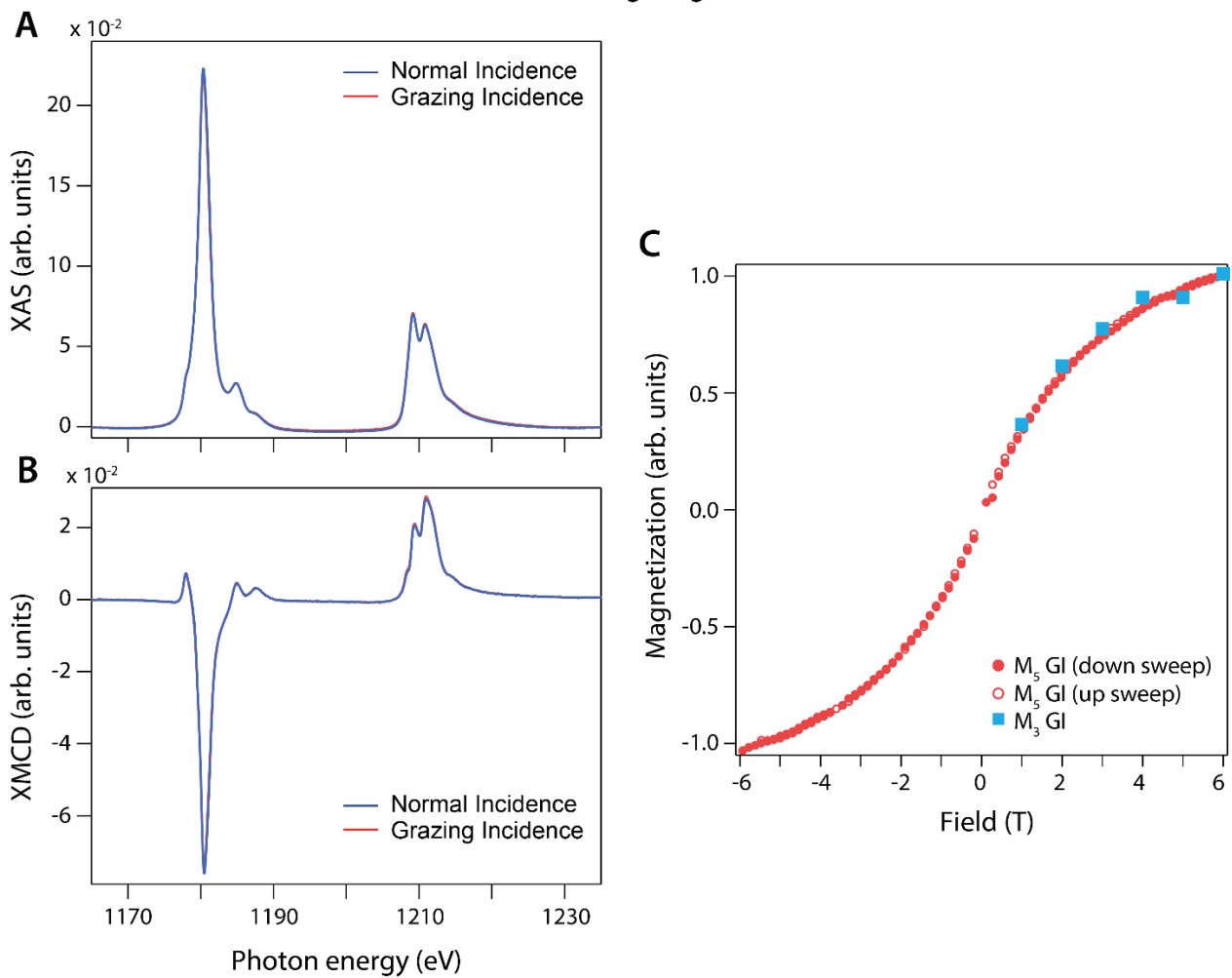

*Figure S3: Anisotropy of Gd clusters on MgO/Ag(100). A: XAS and B: XMCD measurements at the $M_{4,5}$ edge of Gd for two incident angles, normal incidence (NI) and grazing incidence (GI)) at B = 5 T. C: Magnetization loops of the Gd clusters measured at the $M_5$ and $M_3$ edges in GI. All measurement performed at T = 6.5 K.*

### S3. XAS and XMCD of Gd film on MgO/Ag(100)

A gadolinium film of 1.2 ML has been prepared by depositing on a sample with a MgO film grown on Ag(100). The deposition of the Gd layer has been performed by keeping the sample at RT in a base pressure of $1\times10^{-9}$ mbar. The corresponding XAS and XMCD spectra at the $M_{2,3}$ edges are shown in Fig. **S4**. The integrated XMCD signal normalized by the corresponding total XAS is shown in Fig **3F-G** of the main text.

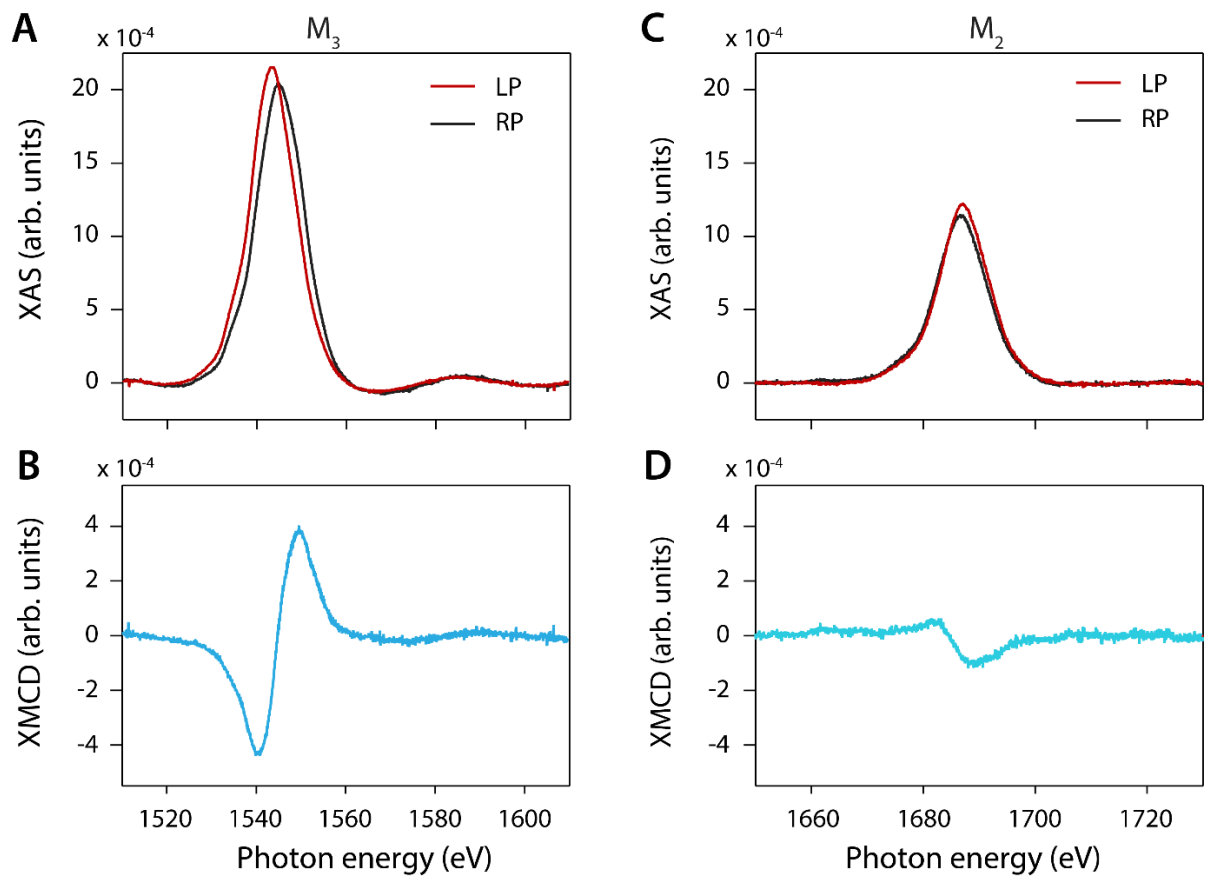

*Figure S4: Spectra at the $M_{2,3}$ edges of Gd film on MgO/Ag(100).* XAS/XMCD measurements performed in total electron yield mode at the **(A-B)** $M_3$ and **(C-D)** $M_2$ edges of Gd. (Gd coverage = 1.2 ML, T = 2.5 K, B = 6.8 T).

## S4. Reference measurements on Gd(III) sulfate octahydrate

Due to its well-known chemical valence and paramagnetic behavior down to low temperatures, Gd(III) sulfate octahydrate is a suitable reference sample to test $M_{2,3}$ transitions in a system with depleted valence *5d*, *6s* and *6p* orbitals *(45)*. The XAS and XMCD spectra of this material were acquired at the 6A MeXiM beamline at Pohang Light Source-II. The sample was prepared by dissolving the Gd salts in pure water, drop casting the solution on the bare copper sample plate, and inserted in the measurement chamber. For that experimental setup, comparative measurements on DySc$_2$N@C$_{82}$ fullerenes powder *(2)* indicate a base temperature of 15 K at the sample plate. Spectra at the $M_{2,3}$ edges acquired at 6.0 T (Fig. **S5**) show line shape features that are very similar to those of Gd atoms and clusters on MgO/Ag(100), see Figs. **2** and **3** of the main text. The $M_3$ edge shows a pronounced peak derivative shape, while a shallow dip is observed at the $M_2$.

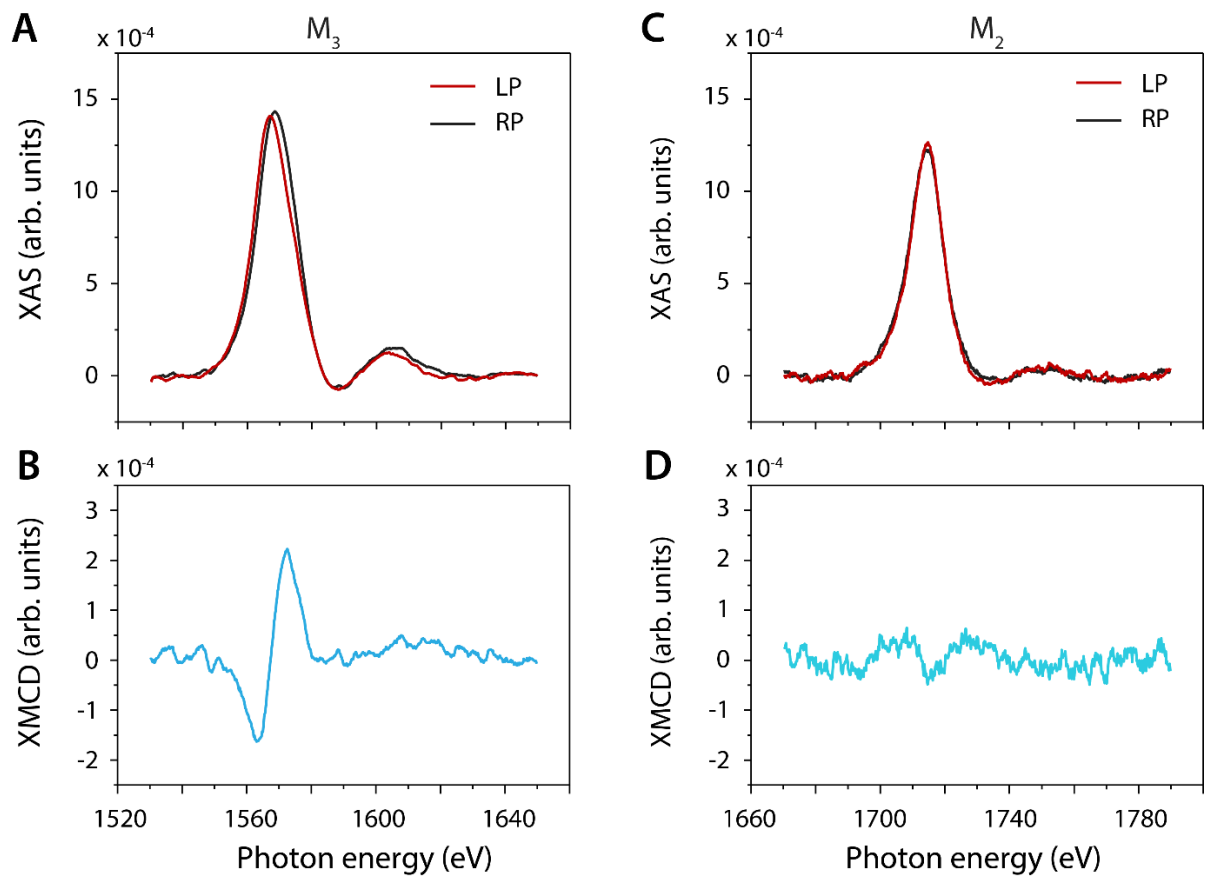

*Figure S5: Reference XAS and XMCD spectra of Gd(III) sulfate octahydrate.* XAS/XMCD measurements performed in total electron yield mode at the **(A-B)** $M_3$ and **(C-D)** $M_2$ edges of Gd (T = 15 K, B = 6.0 T).

## S5. Details of DFT calculations

### A. Orbital-resolved charge analysis of the Gd atoms

Additional insight into the orbital hybridization can be obtained by analyzing the projected density of states (PDOS) for the Gd, O and Mg orbitals shown in Fig. **S6**. Due to the very weak hybridization occurring on the 4f electrons, these states are not presented in those plots. For both Gd/MgO and Gd/MgO/Ag(100) we observe a set of well separated states in the region around the Fermi level. We observe a spin split pair of states indicated with a violet dashed box in Fig. **S6 B** and **D** to have dominant 6s-character. Due to the large radial function of the 6s orbitals, this state is strongly hybridized with p states of the neighboring Mg atoms, see Figs. **S6 A** and **C**. Additional on-site hybridization provides a slight $d_{z^2}$ and $p_z$ character to the states, with a related occupation of about 0.21 and 0.09 electrons, respectively, both with very small net spin polarization. Hence, we expect these states to give negligible contribution to the $M_{1,2,3}$ XMCD edges. The multi-orbital character of this state deforms the original 6s spherical symmetry, as shown in the charge density plot of Fig. **S7 A** and **C**.

The state indicated with a black dashed box is also particularly relevant, as its occupation strongly depends on the presence/absence of the underneath Ag substrate. In both systems, this state has a strong atomic-like $5d_{x^2-y^2}$ character. We find only little hybridization with the Mg p orbitals of the neighboring atoms towards which the 5d orbital lobes are oriented, see also Fig. **S7 B** and **D**. Due to its almost full spin polarization, this state is expected to give a large contribution to the XMCD signal of the $M_{2,3}$ edges. In addition, the localized nature of this state supports the use of multiplet analysis to interpret the XAS and XMCD spectra and infer the depletion.

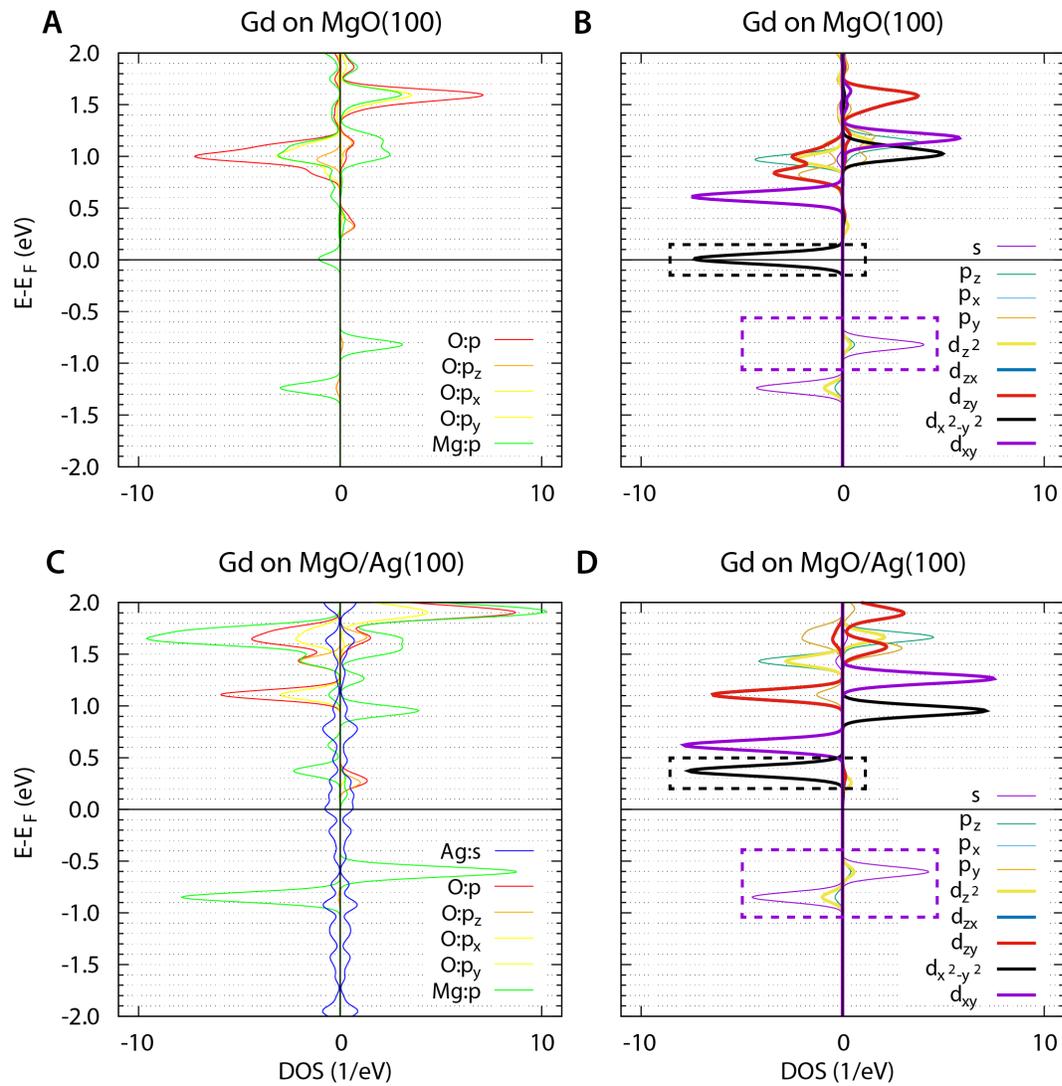

*Figure S6: Orbital analysis of the Gd adatoms. **A**: PDOS of O and Mg orbitals for Gd on MgO(100). **B**: PDOS of 6s, 6p, and 5d orbitals of Gd on MgO(100). **C** and **D**: PDOS for the Gd on MgO/Ag(100) system. All plots are broadened for visual purposes with a Gaussian of approximately 0.1 eV*

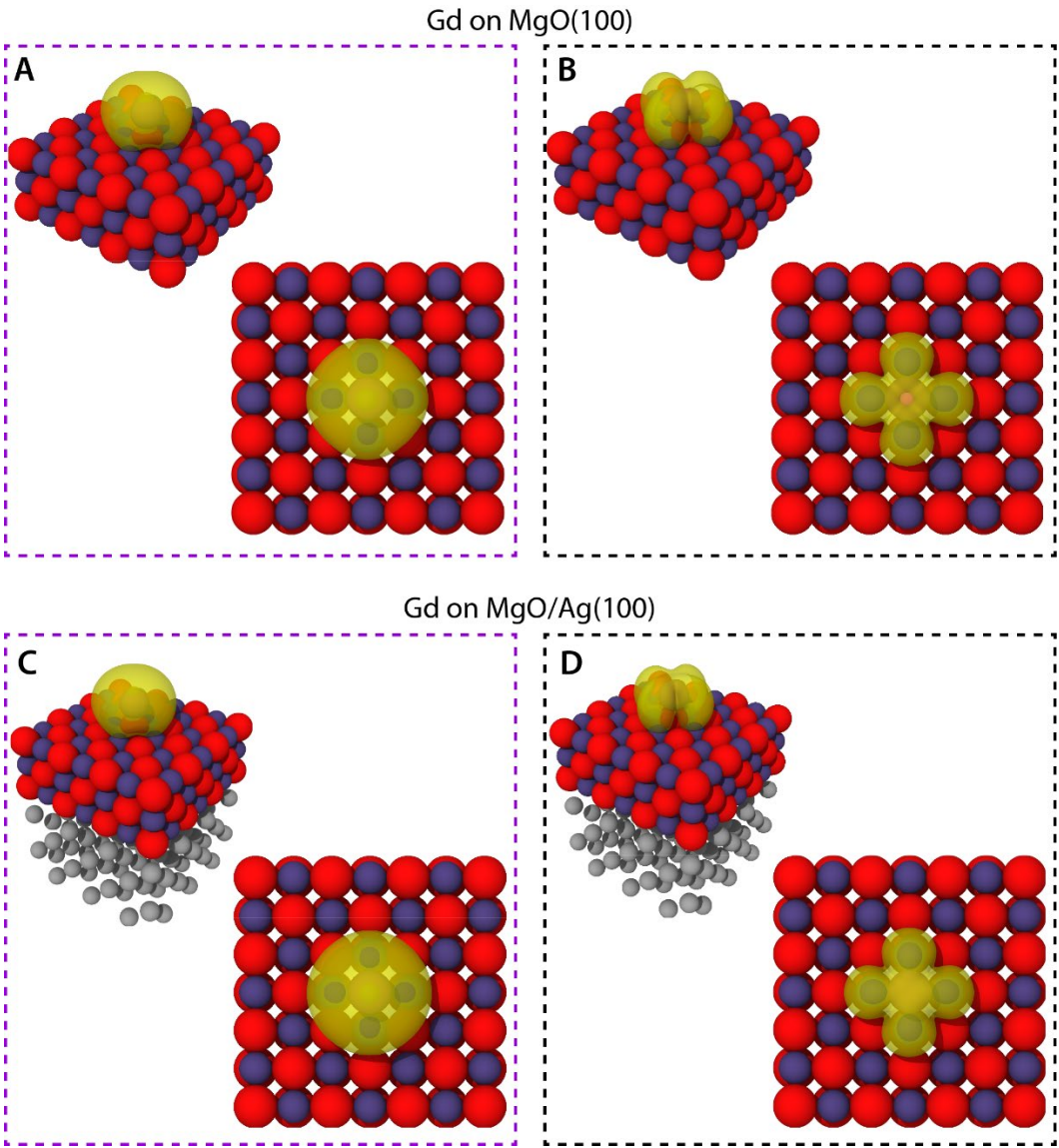

*Figure S7: Integrated densities of Gd states close to Fermi energy. A: Side and top view of states with 6s – like character for the Gd on MgO(100) (purple dashed box of Fig. S6 B). B: Side and top view of states with $5d_{x^2-y^2}$ character for the Gd on MgO(100) (black dashed box of Fig. S6 B). C and D: 6s-like $5d_{x^2-y^2}$-like states for the Gd on MgO/Ag(100) system obtain by plotting the states in purple and black dashed boxes, respectively, indicated in Fig. S6 D. In the plots, red, violet and grey spheres represent O, Mg, and Ag atoms, respectively, whereas the yellow contours are electron states isosurfaces of the indicated PDOS with a value of 0.001.*

### B. Evaluation of crystal field parameters from DFT calculations

We tested two complementary approaches to extract the crystal field (CF) splitting: (i) by diagonalization of the full DFT Hamiltonian and (ii) by construction of maximally localized Wannier functions (MLWFs) of a non-spin-polarized DFT calculation *(46)*. It has been observed that the former will yield a larger CF splitting due to covalent contribution to the total energy, whereas the latter is reduced by about a factor of 0.7 *(47-48)*. For the Wannierization procedure we included projections for Gd:p and d and O:p orbitals. Disentanglement was used to extract isolated bands within a window of 3 eV of the Gd:5d bands. All calculations were performed with the Wannier90 code *(49)*. We found that both approaches yield CFs that result in comparable simulated spectra as the results of the multiplet calculations are not very sensitive to the CF splitting of the empty $5d^0$ orbitals. This, conversely, also means that systematic underestimation of CF splitting for empty states as can be expected from approximate DFT due to the self-interaction error are not expected to play a role here. The crystal field parameters are given in **Table S1**. Additional calculations of Mg-substitutional Gd in bulk MgO indicate that the $e_g$-$t_{2g}$ splitting amounts to 6.4 (4.5) eV for DFT (MLWFs). Hence, the CF splitting for the adatom is drastically reduced compared to bulk case.

*Table S1: CF splitting parameters for the 5d orbitals of Gd on MgO/Ag(100). Given are the dominant orbital character and the energy obtained directly from DFT or by MLWFs. For the former, the energies are obtained from the peaks of the non spin-polarized PDOS.*

| Orbital | Energy (eV) (DFT) | Energy (eV) (MLWFs) |
|---|---|---|
| $d_{z^2}$ | 0.268 | 0.182 |
| $d_{zx}$ | 0.318 | 0.220 |
| $d_{zy}$ | 0.318 | 0.220 |
| $d_{x^2-y^2}$ | -0.602 | -0.418 |
| $d_{xy}$ | -0.282 | -0.197 |
| **Total split:** | **0.92** | **0.64** |

### S6. Additional data for Ho atoms on MgO/Ag(100)

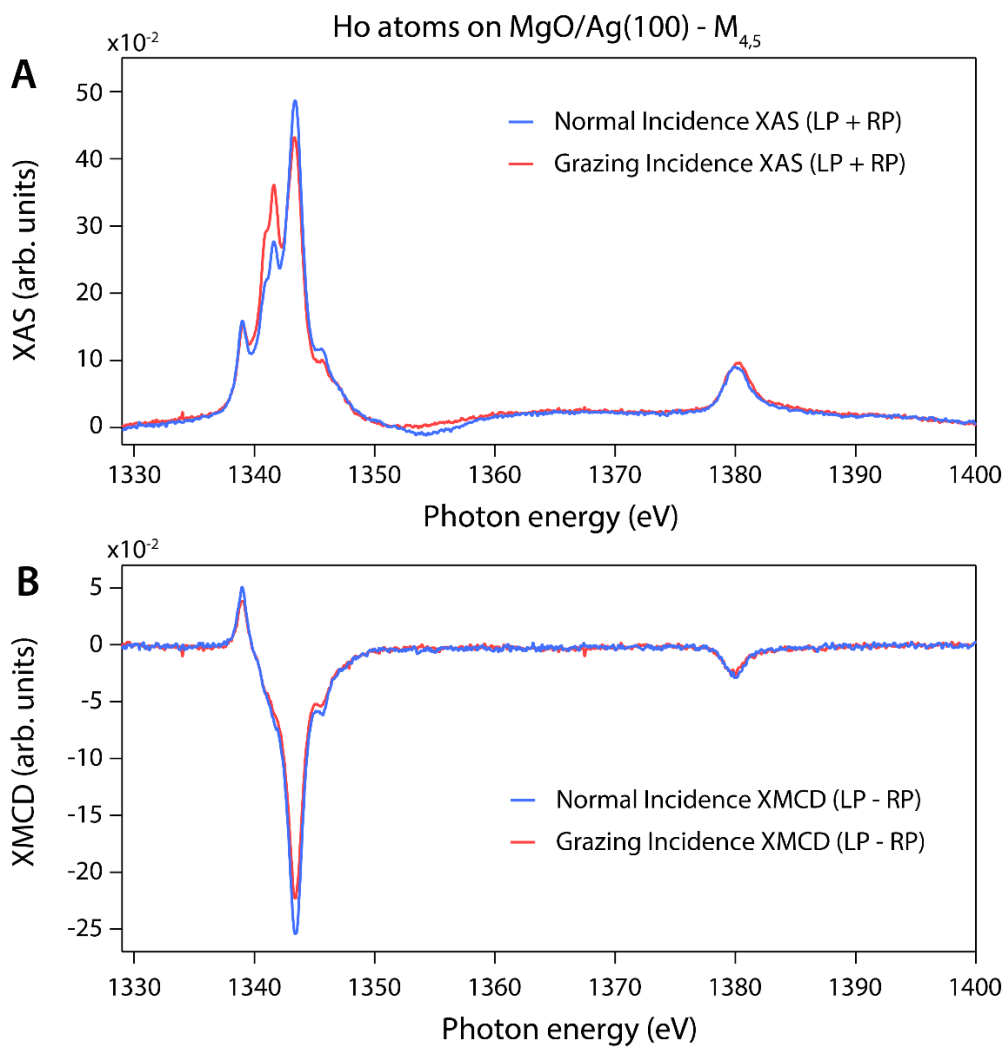

*Figure S8: X-ray spectra at the $M_{4,5}$ edges of Ho atoms on MgO/Ag(100). A: Sum of left (LP) and right polarization (RP) XAS, and B: XMCD signals of Ho atoms at the two indicated incidence geometries (B= 6 T, T =6.5 K, $\Theta_{Ho}$ = 0.03 ML).*

Measurements of the $M_{5,4}$ edges of 0.03 ML of Ho deposited on MgO/Ag(100) are shown in figure **S8**. The line shape and anisotropy at these edges compares very well with previous data *(3,15)*. Minor discrepancies can be attributed to the slightly larger Ho coverage used in this experiment, which may have triggered some amount of cluster nucleation on the surface.

## S7. Parameters for multiplet calculations

As discussed in the Method section of the main text, the Slater integrals to model the electron-electron interaction have been computed with the Cowan's atomic structure code *(40)* and rescaled to fit the experimental spectra. For Gd, all parameters have been rescaled to 0.7 to fit the $M_{4,5}$ spectrum, while an unscaled value of the 3p-4f integral is required to match the amplitude of the XMCD signal at the $M_3$ edge. As discussed in the text, this parameter indirectly couples the 4f to the 5d through a 4f-3p-5d exchange. Due to this coupling, a finite positive value of the XMCD integral at the $M_3$ edge is predicted even in absence of 5d electrons. For Ho, all parameters have been rescaled according to previous values *(50)*. Table S2 summarizes the values used for Gd and Ho. These values are in line with previous results obtained for lanthanide atoms on surfaces *(2,22,50,51)*.

*Table S2: Rescaling parameters for the Slater integrals for multiplet calculations. The coefficients rescale the computed values from the Cowan's atomic structure code (37).*

| Element | 4f-4f | 5d-5d | 6p-6p | 4f-5d | 4f-6p | 3d-4f | 3p-4f | 3p-5d | 3s-4f | 3s-6p |
|---|---|---|---|---|---|---|---|---|---|---|
| Gd | 0.70 | 0.70 | 0.70 | 0.70 | 0.70 | 0.70 | 1.00 | 0.70 | 0.70 | 0.70 |
| Ho | 0.85 | 0.85 | 0.85 | 0.85 | 0.85 | 0.85 | 0.85 | 0.85 | 0.85 | 0.85 |

The crystal field parameters for the 4f electrons have been computed from a point charge electrostatic model *(40)* using the position and values of the neighboring atoms obtained from DFT (see Table S3). The crystal field parameters in the Wybourne notation $\tilde{A}_m^n$ are shown in Tables S4 and S5 for Gd and Ho, respectively. However, the expectation values $\langle r^n \rangle$ of the radial functions from the Cowan's code used to compute those values are known to underestimate the actual splitting observed in experiments *(41)*. Therefore, the crystal field parameters value has been rescaled to match the level splitting computed from a benchmark multiplet code (multiX) *(41)*. The rescaling values $\kappa_m^n$ as well as the final values used in the calculations $A_m^n = \kappa_m^n \tilde{A}_m^n$ are summarized in Tables S4 and S5.

The crystal field parameters for the Gd 5d and 6p electrons have been computed from the energy of the 5d and 6p orbitals, respectively, calculated with DFT, as previously described in Section S5. The related values are summarized in Table S6. Identical values have been used to simulate the $M_{23}$ edges of Ho.

*Table S3: Charges and positions of the neighboring ions included in the multiplet caclulations. The values have been computed from DFT for Gd atom on MgO.*

| Ion | Charge | $d_\perp$ | $d_\parallel$ |
|---|---|---|---|
| O (underneath) | -2e | -215 pm | 0 pm |
| Mg (2nd nearest neighbors) | +2e | -270 pm | 208 pm |
| O (3rd nearest neighbors) | -2e | -270 pm | 294 pm |

*Table S4: Crystal field parameters for the 4f electrons of Gd atoms on MgO/Ag(100). Values from the point charge model $\tilde{A}_m^n$, rescaling coefficients $\kappa_m^n$, and final parameters $A_m^n$ are reported.*

| | $\tilde{A}_0^2$ | $\tilde{A}_0^4$ | $\tilde{A}_4^4$ | $\tilde{A}_0^6$ | $\tilde{A}_4^6$ |
|---|---|---|---|---|---|
| | 486.3 meV | 102.4 meV | 5.7 meV | 32.7 meV | 2.1 meV |
| Gd 4f | $\kappa_0^2$ | $\kappa_0^4$ | $\kappa_4^4$ | $\kappa_0^6$ | $\kappa_4^6$ |
| | $\sqrt{2}$ | $2\sqrt{2}$ | $2\sqrt{2}$ | 4 | 4 |
| | $A_0^2$ | $A_0^4$ | $A_4^4$ | $A_0^6$ | $A_4^6$ |
| | 687.8 meV | 289.7 meV | 16.1 meV | 130.7 meV | 8.3 meV |

*Table S5: Crystal field parameters for the 4f electrons of Ho atoms on MgO/Ag(100). Values from the point charge model $\tilde{A}_m^n$, rescaling coefficients $\kappa_m^n$, and final parameters $A_m^n$ are reported.*

| | $\tilde{A}_0^2$ | $\tilde{A}_0^4$ | $\tilde{A}_4^4$ | $\tilde{A}_0^6$ | $\tilde{A}_4^6$ |
|---|---|---|---|---|---|
| | 410.0 meV | 72.2 meV | 4.0 meV | 18.8 meV | 1.2 meV |
| Ho 4f | $\kappa_0^2$ | $\kappa_0^4$ | $\kappa_4^4$ | $\kappa_0^6$ | $\kappa_4^6$ |
| | $\sqrt{2}$ | $2\sqrt{2}$ | $2\sqrt{2}$ | 4 | 4 |
| | $A_0^2$ | $A_0^4$ | $A_4^4$ | $A_0^6$ | $A_4^6$ |
| | 579.8 meV | 204.3 meV | 11.3 meV | 75.2 meV | 4.8 meV |

*Table S6: Crystal field parameters for the 5d and 6p electrons of Gd atoms on MgO/Ag(100). Values have been computed from the Wannierization procedure described in section S5. Identical values have been used for Ho.*

| Gd 5d | $A_0^2$ | $A_0^4$ | $A_4^4$ | Gd 6p | $A_0^2$ |
|---|---|---|---|---|---|
| | 1470 meV | -546 meV | -401 meV | | 283 meV |